\documentclass[12pt]{article} \textheight=23cm \textwidth=16cm
\usepackage{graphicx}
\usepackage{subfigure}

\def\beq{\begin{equation}}
\def\eeq{\end{equation}}
\def\barr#1{\begin{array}{#1}}
\def\earr{\end{array}}
\def\beqar{\begin{eqnarray}}
\def\eeqar{\end{eqnarray}}
\def\beqars{\begin{eqnarray*}}
\def\eeqars{\end{eqnarray*}}
\def\bitem{\begin{itemize}}
\def\eitem{\end{itemize}}

\def\bc{\begin{center}}
\def\ec{\end{center}}
\def\beq{\begin{equation}}
\def\eeq{\end{equation}}
\def\bea{\begin{eqnarray}}
\def\eea{\end{eqnarray}}
\def\bit{\begin{itemize}}
\def\eit{\end{itemize}}
\def\ben{\begin{enumerate}}
\def\een{\end{enumerate}}
\def\ba{\begin{array}}
\def\ea{\end{array}}
\def\bc{\begin{center}}
\def\ec{\end{center}}

\def\dReg(#1,#2){\SetOffset(#1,#2)\BCirc(0,0){4}
    \Line(2.8,2.8)(-2.8,-2.8)
    \Line(2.8,-2.8)(-2.8,2.8)\SetOffset(0,0)}

\newenvironment{rowvec2}{\left(\begin{array}{cc}}{\end{array}\right)}
\newenvironment{colvec}{\left(\begin{array}{c}}{\end{array}\right)}
\def\bcol{\begin{colvec}}
\def\ecol{\end{colvec}}
\def\brow2{\begin{rowvec2}}
\def\erow2{\end{rowvec2}}

\begin{document}



\catcode`\@=11
\def\lesssim{\mathrel{\mathpalette\vereq<}}
\def\gtrsim{\mathrel{\mathpalette\vereq>}}
\def\vereq#1#2{\lower3pt\vbox{\baselineskip0pt \lineskip0pt
\ialign{$\m@th#1\hfill##\hfill$\crcr#2\crcr\sim\crcr}}}
\catcode`\@=12

\def\alt{\lesssim}
\def\agt{\gtrsim}
\bigskip

\begin{center}
{\Large \bf $B^0 - \overline{B^0}$ mixing, $B\rightarrow J/\psi K_s$ and 
$B\rightarrow X_d ~\gamma$ \\ in general MSSM
\footnote{
Talk presented by P. Ko at SUSY02, Hamburg, June 2002, 
to appear in Proceedings,
}}

\bigskip

P. Ko$^{a,b}$, ~~~G. Kramer$^c$, ~~~Jae-hyeon Park$^a$ 

\bigskip

$^a$ Department of Physics, KAIST \\ Daejeon 305-701, Korea
\bigskip

$^b$ 
MCTP and Randall Lab, 
University of Michigan
\\
Ann Arbor, MI 48109, U.S.A.
\bigskip

$^c$  II. Institut f\"{u}r Theoretische Physik, 
Universit\"{a}t Hamburg \\ D-22761 Hamburg, Germany

\end{center}






\begin{abstract}
We consider the gluino-mediated SUSY contributions to $B^0 - \overline{B^0}$ 
mixing, $B\rightarrow J/\psi K_s$ and $B\rightarrow X_d \gamma$ in the mass
insertion approximation.
We find the $(LL)$ mixing parameter can be as large as 
$| (\delta_{13}^d)_{LL} | \lesssim 2 \times 10^{-1}$, 
but the $(LR)$ mixing is strongly constrained by the $B\rightarrow X_d \gamma$ 
branching ratio and we find $| (\delta_{13}^d)_{LR} | \lesssim 10^{-2}$.
The implications for the direct CP asymmetry in $B\rightarrow X_d \gamma$ 
and the dilepton charge asymmetry ($A_{ll}$) are also discussed, where 
substantial deviations from the standard model (SM) predictions are possible.
\end{abstract}






Recent observations of large CP violation in $B\rightarrow J/\psi K_s$
\cite{exp:sin2beta,masiero2002} giving 
\begin{equation}
\sin 2 \beta = ( 0.79 \pm 0.10 )
\end{equation}
confirm the SM prediction and begin to put a strong constraint on new physics
contributions to $B^0 - \overline{B^0}$ mixing and $B\rightarrow J/\psi K_s$,
when combined with $\Delta m_{B_d} = (0.472 \pm 0.017 )~{\rm ps}^{-1}$ 
\cite{pdg}. Since the decay $B\rightarrow J/\psi K_s$ is dominated by the 
tree level SM process $b\rightarrow c \bar{c} s$, we expect the new physics 
contribution may affect significantly only the $B^0 - \overline{B^0}$ mixing
and not the decay $B\rightarrow J/\psi K_s$. 
However, in the presence of new physics contributions to 
$B^0 - \overline{B^0}$ mixing, the same new physics would generically affect 
the $B\rightarrow X_d \gamma$ process. 
%
In this talk, we 
present our recent work on 
$B^0 -  \overline{B^0}$ mixing, $B\rightarrow J/\psi K_s$ and 
$B_d \rightarrow X_d \gamma$ in general SUSY models where flavor and CP 
violation due to the gluino mediation can be important \cite{kkp}, where
more detailed references can be found. 
We use the mass insertion approximation (MIA) for this purpose.
Comprehensive work has been done for the first two observables in the MIA
considering $\Delta m_{B_d}$ and $\sin2\beta$ constraints only (see Ref.~
\cite{masiero2002}  for the most recent studies with such an approach). 
In our work, 
we also include the dilepton charge asymmetry $A_{ll}$ and the 
$B_d \rightarrow X_d \gamma$ branching ratio constraint extracted from the 
recent experimental 
upper limit on the $B\rightarrow \rho\gamma$ branching ratio \cite{b2rho}
$B( B \rightarrow \rho \gamma ) < 2.3 \times 10^{-6}, $ and rederive
the upper limits on the $(\delta_{13}^d )_{LL}$ and $(\delta_{13}^d )_{LR}$ 
mixing parameters assuming that only one of these gives a dominant SUSY 
contribution in addition to the standard model (SM) contribution. 
In addition we study the direct CP asymmetry in $B_d \rightarrow X_d \gamma$
on the basis of our result for the SUSY contribution, and 
discuss how much deviations from the SM predictions are expected. 
Although we confine ourselves here to the gluino-mediated SUSY constributions 
only, our strategy can be extended to any new physics scenario
with a substantial constribution to $B^0 - \overline{B^0}$ mixing and 
$B\rightarrow X_d \gamma$. 



The effective Hamiltonian for $B^0 -\overline{B^0}$ mixing
($\Delta B = 2$) and $B\rightarrow X_d \gamma$ including the gluino 
loop contributions can be found in Ref.~\cite{masiero2002} and 
Ref.~\cite{kkp}, respectively.
The $\Delta B=2$ effective Hamiltonian will contribute to $\Delta m_B$,
the dilepton charge asymmetry and the time dependent CP asymmetry in the decay
$B\rightarrow J/\psi K_s$ via the phase of the $B^0 - \overline{B^0}$ mixing.
Defining the mixing matrix element by
\begin{equation}
M_{12} (B^0) \equiv {1\over 2 m_B}~\langle B^0 | H_{\rm eff}^{\Delta B = 2}
| \overline{B^0} \rangle \,  
\end{equation}
one has $\Delta m_{B_d} = 2 | M_{12} (B_d^0) |$. 
On the other hand, the phase of the $B^0 - \overline{B^0}$ mixing amplitude 
$M_{12} (B^0) \equiv \exp( 2 i \beta^{'} ) ~| M_{12} (B^0) |$
appears in the time dependent asymmetry : 
$A_{\rm CP}^{\rm mix} ( B^0 \rightarrow J/\psi K_s )= \sin 2 \beta^{'}
~\sin \Delta m_{B_d}t$.
%
%
Finally, the dilepton charge asymmetry $A_{ll}$ 
is also determined by $M_{12} (B^0)$, albeit a possible long distance 
contribution to $\Gamma_{SM} ( B^0 )$:
\begin{equation}
\label{eq:all}
A_{ll} \equiv {N(BB) - N(\bar{B}\bar{B}) \over N(BB) + N(\bar{B}\bar{B})}
\approx {\rm Im} \left( \Gamma_{12} / M_{12} \right) 
\approx 
{\rm Im} \left( { \Gamma_{12}^{\rm SM} \over
M_{12}^{\rm SM} + M_{12}^{\rm SUSY} } \right).
\end{equation}
Here $M_{12}, \Gamma_{12}$ are the matrix elements of the Hamiltonian in the
$(B^0, \overline{B^0} )$ basis:
\[
{1\over 2 m_B}~\langle \overline{B} | H_{\rm full} | B \rangle = 
M_{12} - {i \over 2} \Gamma_{12}.
\]
We have used the fact $\Gamma_{12}^{\rm FULL} \approx \Gamma_{12}^{\rm SM}$.
The SM prediction is 
$-1.54 \times 10^{-3} \leq A_{ll}^{\rm SM} \leq -0.16 \times 10^{-3}$,  
whereas the current world average is \cite{nir}
$A_{ll}^{\rm exp} \approx (0.2 \pm 1.4) \times 10^{-2}$.




The effective Hamiltonian relevant to $\Delta B = 1$ processes involves 
four quark operators and $b\rightarrow d \gamma$ and $b\rightarrow d g$ 
penguin operators. 
Since we are not going to discuss $\Delta B =1$ nonleptonic decays due to
theoretical uncertainties related with factorization, we shall consider
the inclusive radiative decay $B\rightarrow X_d \gamma$ only. 
The relevant effective Hamiltonian for this process is given by \cite{ali}. 
Varying $f_{B_d}$, $|V_{ub}|$, and $|V_{cb}|$ in the uncertainty range, 
and $\gamma$ between $(54.8 \pm 6.2)^\circ$ \cite{Ciuchini:2000de}, 
we get the branching ratio for this decay in the SM to be
$8.9 \times 10^{-6} - 1.1 \times 10^{-5}$.
The direct CP asymmetry in the SM is about
$-15 \% - -10 \%$ \cite{ali}. We have updated the previous predictions by 
Ali et al. \cite{ali} using the present values of CKM parameters.

\begin{figure}
\centering
\subfigure[$LL$ mixing only]{\raisebox{1.5mm}{\includegraphics[width=7cm,height=7cm]{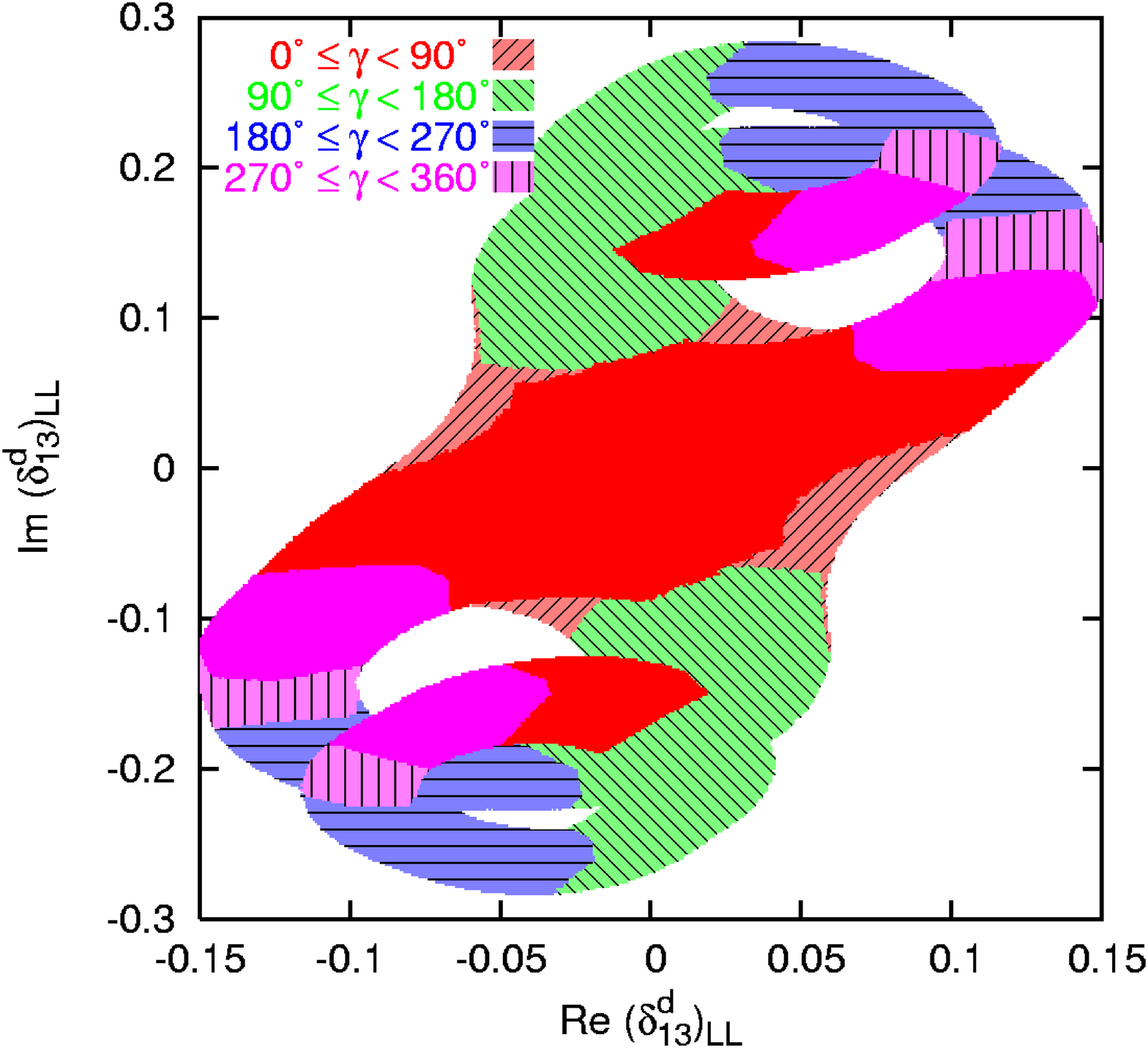}}}
\subfigure[$A_{ll}$]{\includegraphics[width=6.8cm,height=7.06cm]%
{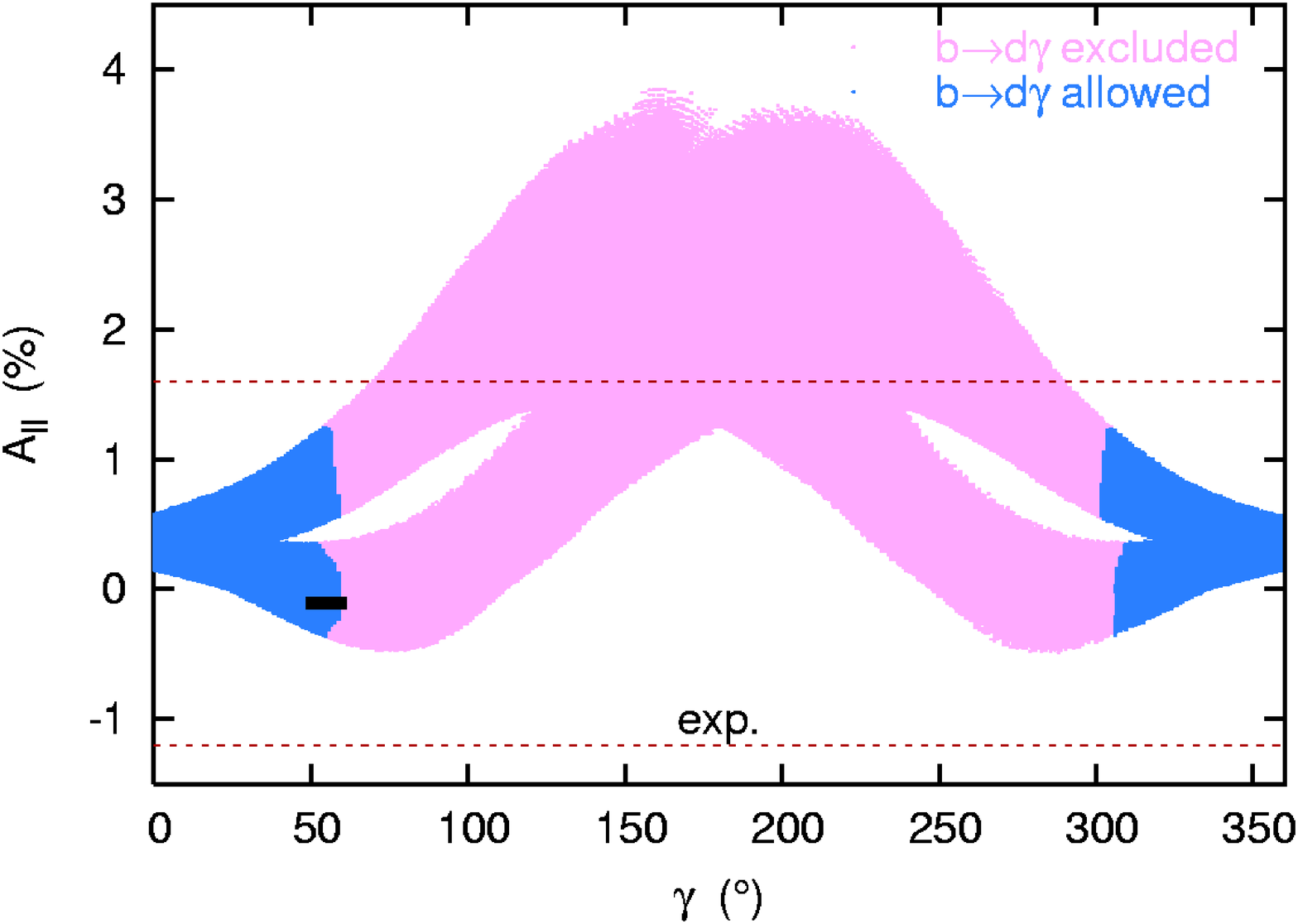}}
\subfigure[B ($B\rightarrow X_d \gamma)$]{\includegraphics[width=7cm,height=7cm]
{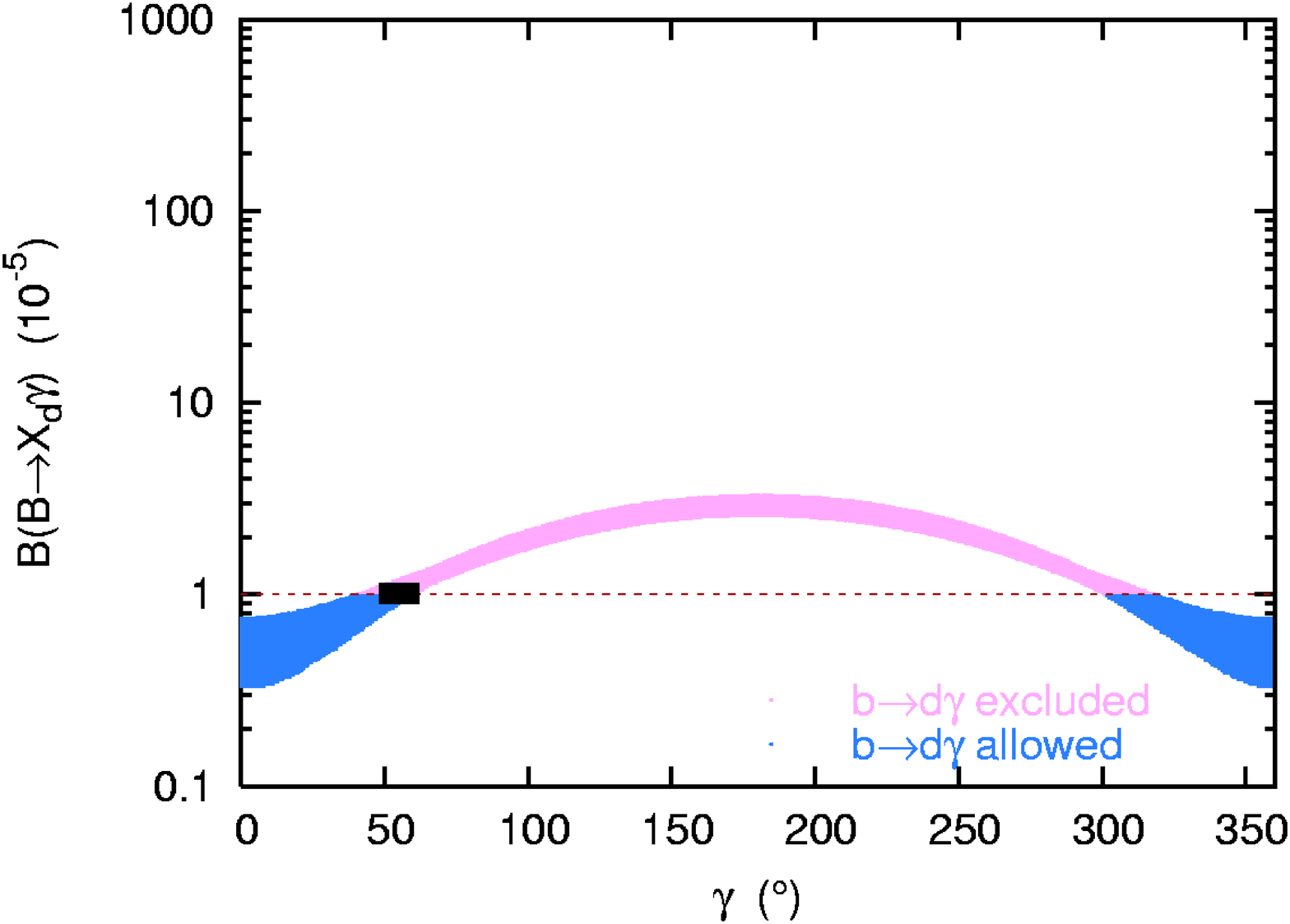}}
\subfigure[$A_{\rm CP}^{b\rightarrow d\gamma}$]{\includegraphics[width=7cm,height=6.89cm]
{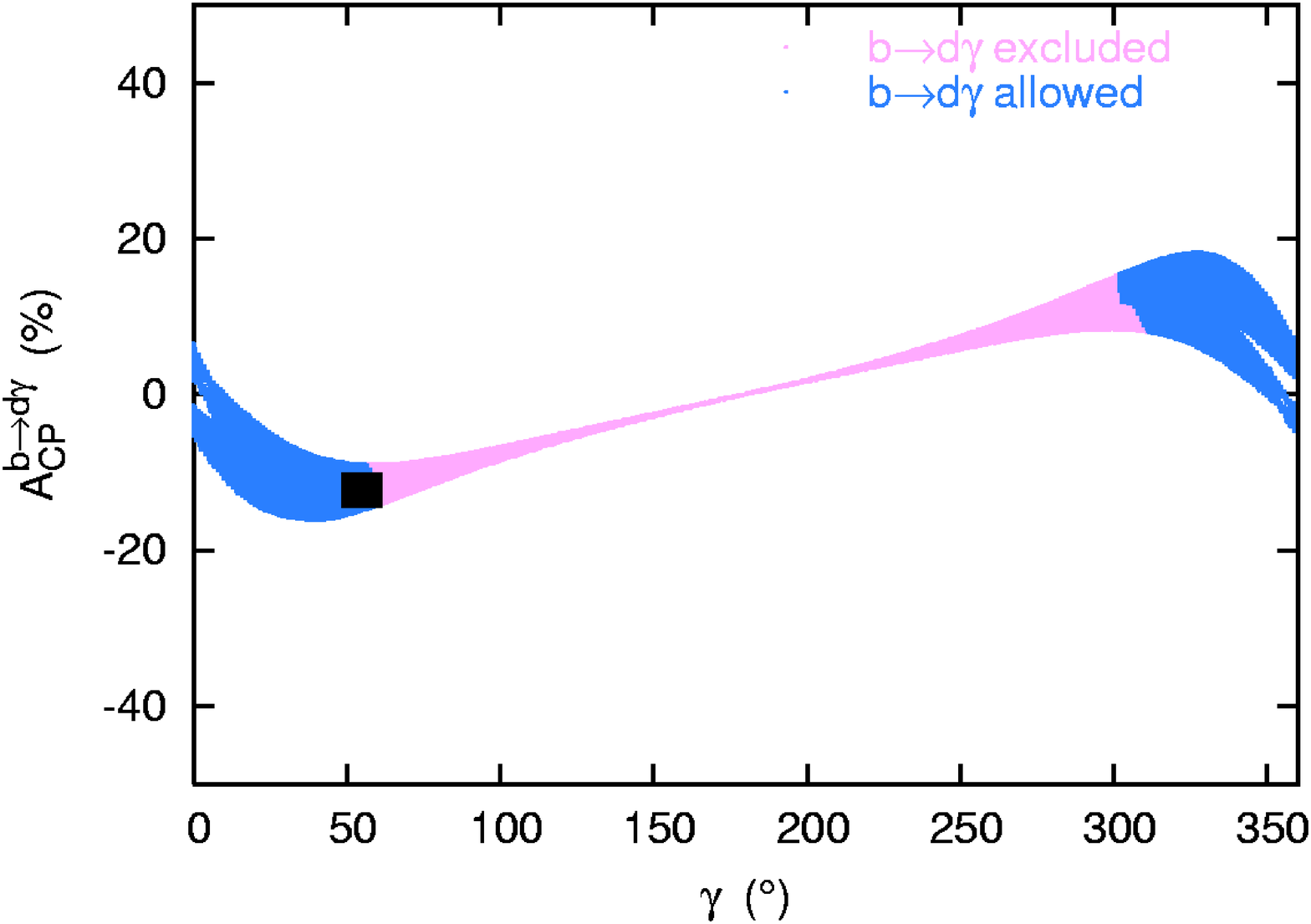}}
\caption{
(a) The allowed range in the $LL$ insertion case for 
the parameters 
$( {\rm Re}(\delta_{13}^d )_{AB}, {\rm Im} (\delta_{13}^d )_{AB})$ for 
different values of the KM angle $\gamma$ with different color codes:
dark (red) for $0^{\circ} \leq \gamma \leq 90^{\circ}$, light gray (green)
for  $90^{\circ} \leq \gamma \leq 180^{\circ}$, very dark (blue) for
$180^{\circ} \leq \gamma \leq 270^{\circ}$ and gray (magenta) for 
$270^{\circ} \leq \gamma \leq 360^{\circ}$.
The region leading to a too large branching ratio for 
$B_d \rightarrow X_d \gamma$ is colored lightly and covered by parallel lines.
(b), 
(c) and (d) are  $A_{ll}$, 
B  ($B\rightarrow X_d \gamma)$ and direct CP asymmetry
therein as functions of KM angle $\gamma$.
}
\label{fig:d13}
\end{figure}%

\begin{figure}
\centering
\subfigure[$LR$ mixing only]{\raisebox{1.5mm}{\includegraphics[width=7.1cm,height=7cm]{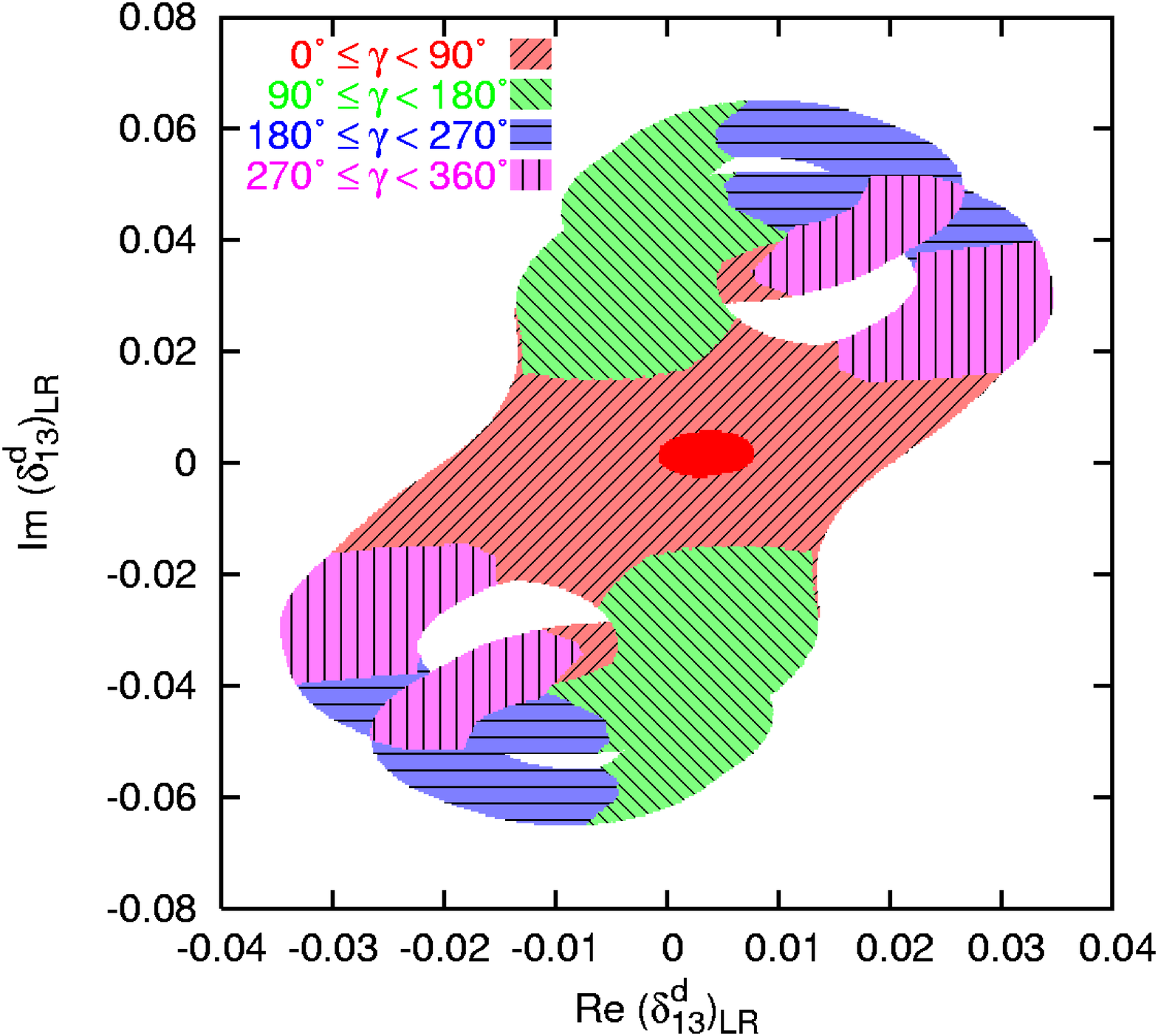}}}
\subfigure[$A_{ll}$]{\includegraphics[width=6.8cm,height=7.06cm]
{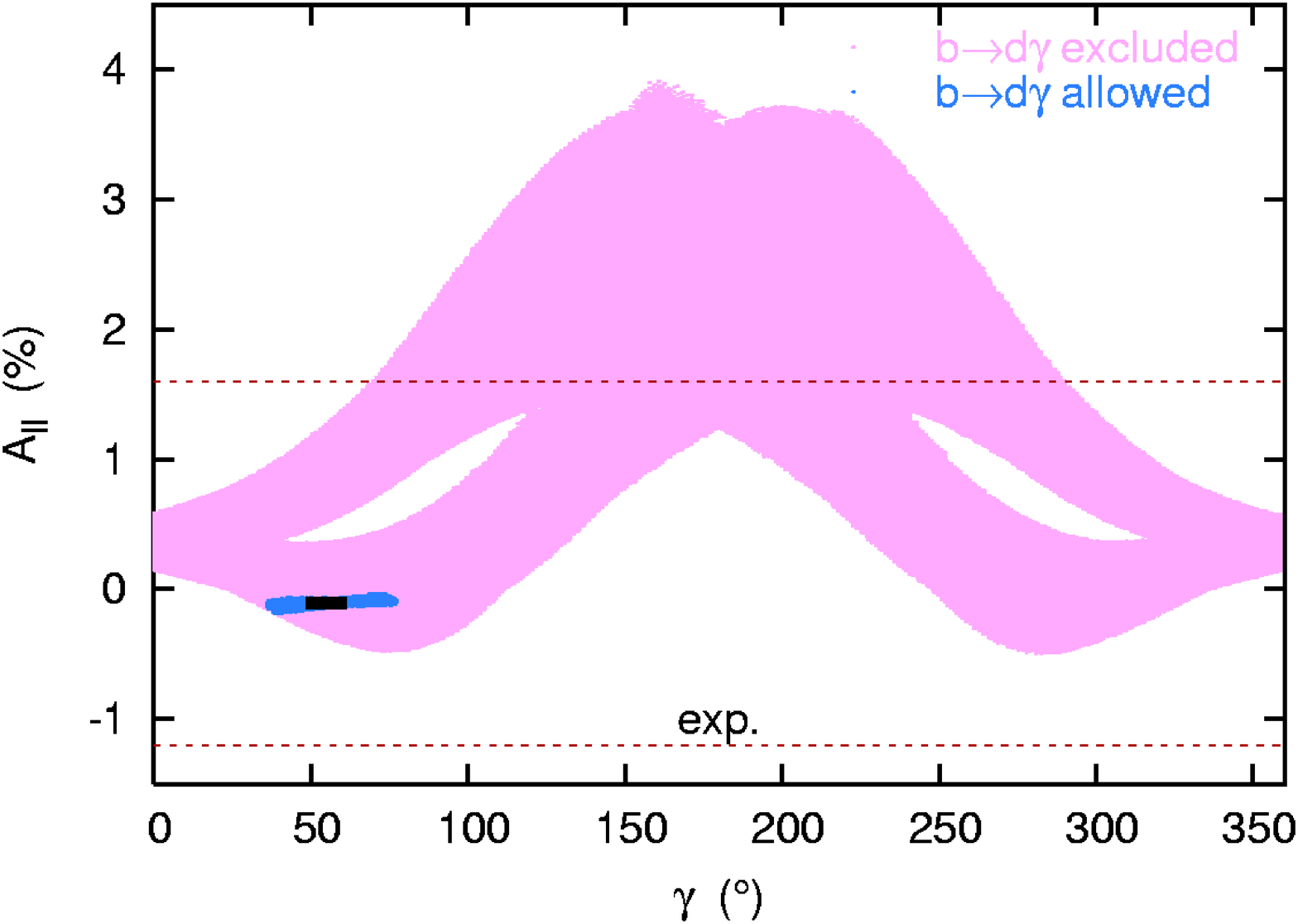}}
\subfigure[B ($B\rightarrow X_d \gamma)$]{\includegraphics[width=6.9cm,height=7cm]
{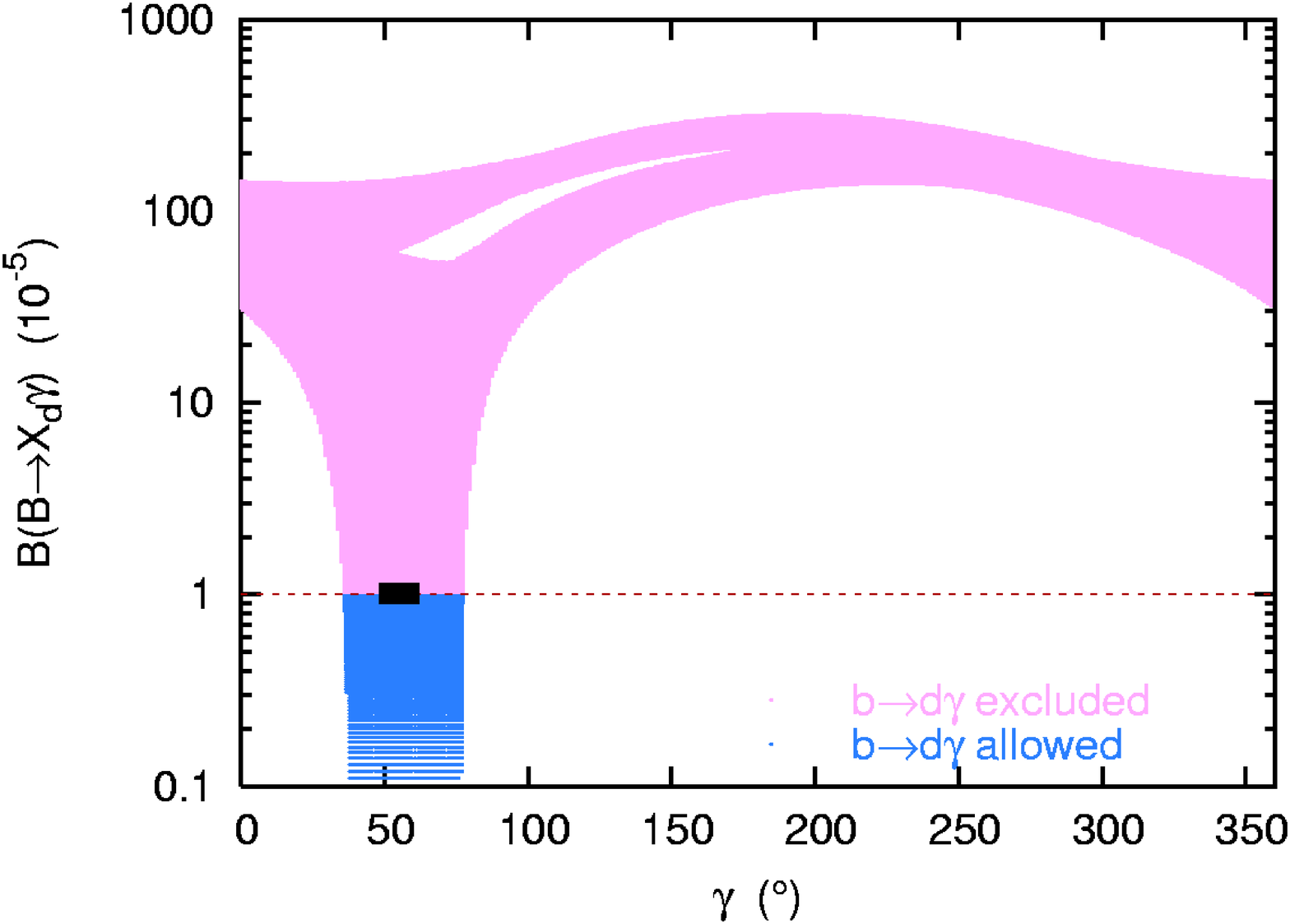}}
\subfigure[$A_{\rm CP}^{b\rightarrow d\gamma}$]{\includegraphics[width=7cm,height=6.89cm]
{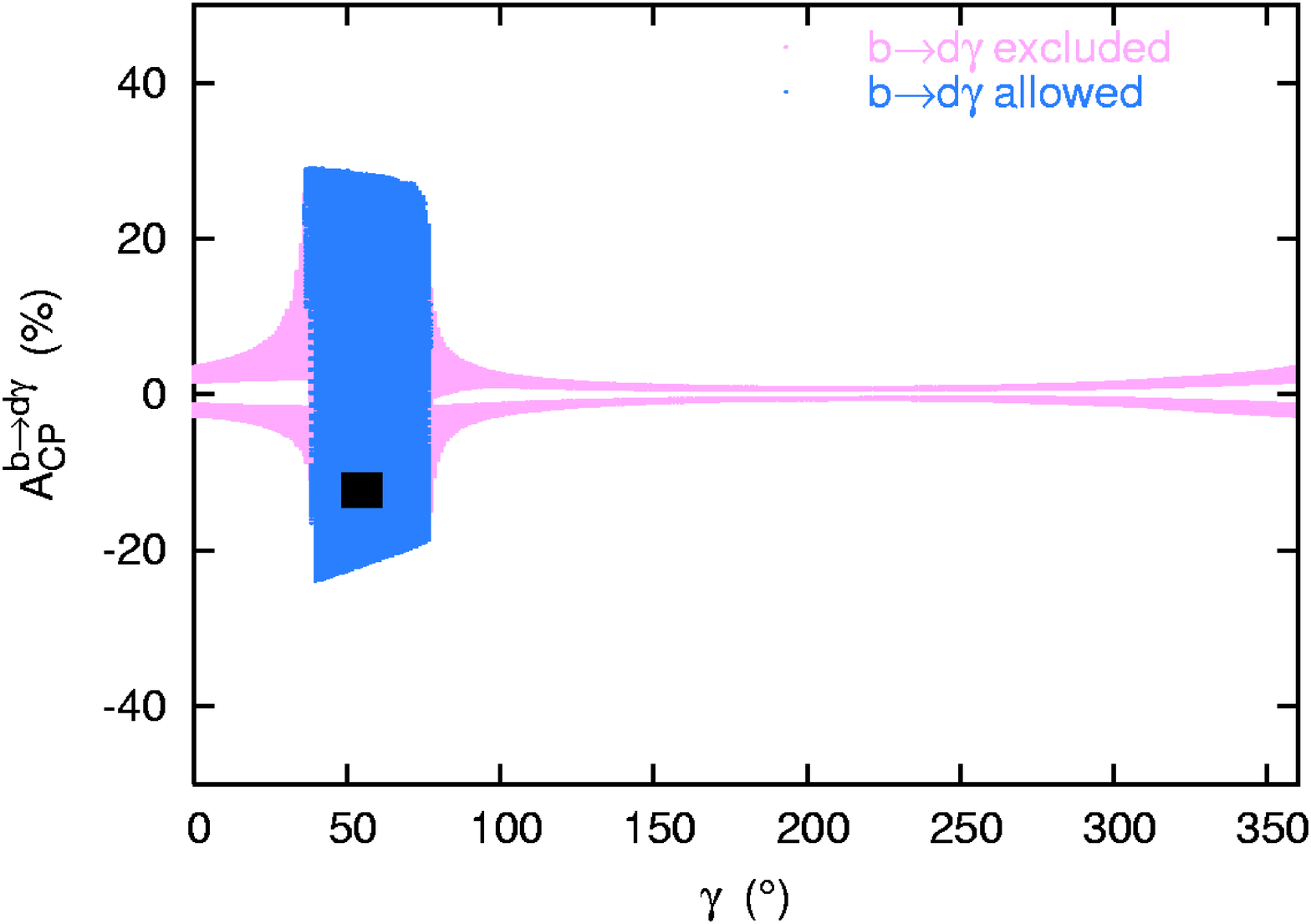}}
\caption{
(a) The allowed range in the $LR$ insertion case for 
the parameters 
$( {\rm Re}(\delta_{13}^d )_{AB}, {\rm Im} (\delta_{13}^d )_{AB})$ for 
different values of the KM angle $\gamma$ with different color codes:
dark (red) for $0^{\circ} \leq \gamma \leq 90^{\circ}$, light gray (green)
for  $90^{\circ} \leq \gamma \leq 180^{\circ}$, very dark (blue) for
$180^{\circ} \leq \gamma \leq 270^{\circ}$ and gray (magenta) for 
$270^{\circ} \leq \gamma \leq 360^{\circ}$.
The region leading to a too large branching ratio for 
$B_d \rightarrow X_d \gamma$ is colored lightly and covered by parallel lines.
(b), 
(c) and (d) are  $A_{ll}$, 
B  ($B\rightarrow X_d \gamma)$ and direct CP asymmetry
therein as functions of KM angle $\gamma$.
}
\label{fig:all}
\end{figure}

In the numerical analysis, we impose the following quantities as constraints :
\[
\Delta m_{B_d} = (0.472 \pm 0.017) ~{\rm ps}^{-1},~~~
A_{\rm CP}^{\rm mix} =  ( 0.79 \pm 0.10 ),~~~
{\rm Br} ( B \rightarrow X_d \gamma ) < 1 \times 10^{-5}
\]
For the dilepton charge asymmetry, 
we do not use this constraint to restrict the allowed parameter space, 
since it is weaker than the other constraints. We 
indicate the parameter space where the resulting $A_{ll}$ falls out of the 
$1\sigma$ range. 
We impose these constraints at 68 \% C.L. ($1 \sigma$) as we vary the KM
angle $\gamma$ between 0 and $2 \pi$. In all cases, we set the common squark 
mass $\tilde{m} = 500$ GeV and $x = 1$ ($m_{\tilde{g}} = \tilde{m}$).
Finally for the mass insertion parameters $(\delta_{13}^d )_{AB}$, we 
consider two cases.

In Figs.~\ref{fig:d13} (a), we show the allowed parameter space in 
the $( {\rm Re}(\delta_{13}^d )_{LL}, {\rm Im} (\delta_{13}^d )_{LL})$ plane
for different values of the KM angle $\gamma$ with different color codes:
dark (red) for $0^{\circ} \leq \gamma \leq 90^{\circ}$, light gray (green)
for  $90^{\circ} \leq \gamma \leq 180^{\circ}$, very dark (blue) for
$180^{\circ} \leq \gamma \leq 270^{\circ}$ and gray (magenta) for 
$270^{\circ} \leq \gamma \leq 360^{\circ}$. The region leading 
to a too large branching ratio for $B_d \rightarrow X_d \gamma$ is covered 
by slanted lines. And the region where $A_{ll}$ falls out of the data within
$1\sigma$ range is already excluded by the $B\rightarrow X_d \gamma$ 
branching ratio constraint [ Fig.~1 (b) ]. 
Note that the KM angle $\gamma$ should be
in the range between  $\sim - 60^\circ$ and $\sim + 60^{\circ}$, and $A_{ll}$
can have the opposite sign compared to the SM prediction, even if the KM 
angle  is the same as its SM value $\gamma \simeq 55^{\circ}$ due to the
SUSY contributions to $B^0 - \overline{B^0}$ mixing.  
This is entirely different from Ref.~\cite{masiero2002}, where the KM 
angle $\gamma$ is not constrained at all. $B\rightarrow X_d \gamma$ plays
an important role here.
In Figs.~\ref{fig:d13} (c) and (d),
we show the branching ratio of $B_d \rightarrow X_d \gamma$ and the direct 
CP asymmetry therein, respectively, as functions of the KM angles $\gamma$ 
for the $LL$ insertion only. The SM predictions 
\[
B (B_d \rightarrow X_d \gamma) = (0.9 - 1.1 ) \times 10^{-5},
~~~A_{\rm CP}^{b\rightarrow d \gamma} = (-15 \sim -10) \%
\] 
are indicated by the black boxes. 
In this case, the KM angle $\gamma$ is constrained in the range 
$\sim - 60^\circ$ and $\sim + 60^{\circ}$.
The direct CP asymmetry is predicted to be between $\sim - 15\%$ and 
$\sim +20\%$.
In the $LL$ mixing case, the SM gives the dominant contribution to  
$B_d \rightarrow X_d \gamma$, but the KM angle can be different from the 
SM case, because SUSY contributions to the $B^0 - \overline{B^0}$ mixing can 
be significant and the preferred value of $\gamma$ can change from the SM 
KM fitting. This is the same in the rare kaon decays and the results 
obtained in Ref.~\cite{ko2} apply without modifications.
If the KM angle $\gamma$ is substantially different from the SM value (say,
$\gamma = 0$), we could anticipate large deviations in the 
$B_d \rightarrow X_d \gamma$ branching ratio and the direct CP violation
thereof. 

For the $LR$ mixing [ Fig.~2 (a) ], 
the $B(B_d \rightarrow X_d \gamma)$ puts an even stronger constraint on 
the $LR$ insertion, whereas the $A_{ll}$ does not play any role. 
In particular, the KM  angle $\gamma$ can not be too much different from 
the SM value in the $LR$ mixing case, once the 
$B(B_d \rightarrow X_d \gamma)$  constraint is included. 
Only $30^{\circ} \lesssim \gamma \lesssim 80^{\circ}$ is compatible with all 
the data from the $B$ system, even if we do not consider the $\epsilon_K$ 
constraint. The resulting parameter space is significantly reduced compared 
to the result obtained in Ref.~\cite{masiero2002}. 
The limit on the $LR$ insertion parameter will become even stronger as the
experimental limit on $B_d \rightarrow X_d \gamma$ will be improved in the 
future. 
In Fig.~2 (b), we show the predictions for $A_{ll}$ as a function of the 
KM angle $\gamma$ for the $LR$ insertion only. On the other hand, for the 
$LR$ insertion case, the $B\rightarrow X_d \gamma$ constraint rules out 
essentially almost all the parameter space region, and the resulting 
$A_{ll}$ is essentially the same as for the SM case. In Figs.~2 (c) and (d),
we show the branching ratio of $B_d \rightarrow X_d \gamma$ and the direct 
CP asymmetry therein, respectively, as functions of the KM angles $\gamma$ 
for the $LR$ insertion only. As  before, the black boxes 
represent the SM predictions for $B (B_d \rightarrow X_d \gamma)$ and 
the direct CP asymmetry therein. In the $LR$ insertion case, there could 
be substantial deviations in both the branching ratio and the CP asymmetry 
from the SM predictions, even if the $\Delta m_B$ and $\sin 2 \beta$ is 
the same as the SM predictions as well as the data. For the $LL$ insertion, 
such a large deviation is possible, since the KM angle $\gamma$ can be 
substantially different from the SM value. On the other hand, for the $LR$ 
mixing, the large deviation comes from the complex $( \delta_{13}^d )_{LR}$ 
even if the KM angle is set to the same value as in the SM. 
The size of $( \delta_{13}^d )_{LR}$ is too small to affect the 
$B^0 - \overline{B^0}$ mixing, but is still large enough too affect 
$B\rightarrow X_d \gamma$. Our model independent study indicates that
the current data on the $\Delta m_B$, $\sin2\beta$ and $A_{ll}$ do still 
allow a possibility for large deviations in $B\rightarrow X_d \gamma$,
both in the branching ratio and the direct CP asymmetry thereof. 
The latter variables are indispensable to test completely the KM paradigm for
CP violation and get ideas on possible new physics with new flavor/CP 
violation in $b\rightarrow d$ transition.


In conclusion, we considered the gluino-mediated SUSY contributions to 
$B^0 - \overline{B^0}$ mixing, $B\rightarrow J/\psi K_s$ and 
$B\rightarrow X_d \gamma$ in the mass insertion approximation.
We find that the $(LL)$ mixing parameter can be as large as 
$| (\delta_{13}^d)_{LL} | \lesssim 2 \times 10^{-1}$, but the 
$(LR)$ mixing is strongly constrained by the $B\rightarrow X_d \gamma$ 
branching ratio: $| (\delta_{13}^d)_{LR} | \lesssim 10^{-2}$.
The implications for the direct CP asymmetry in $B\rightarrow X_d \gamma$ 
are also discussed, where substantial deviations from the SM predictions are 
possible  both in the $LL$ and $LR$ insertion cases for different reasons,
as discussed in the previous paragraphs. 
Our analysis demonstrates that all the observables, $A_{ll}$, the branching 
ratio of $B\rightarrow X_d \gamma$ and the direct CP violation thereof 
are very important, since they could provide informations on new flavor
and CP violation from $(\delta_{13}^d )_{LL,LR}$ (or any other new physics 
scenarios with new flavor/CP violations).
Also they are indispensable in order that we can ultimately test 
the KM paradigm for CP vioaltion in the SM. 

%
%

This work is supported in part by BK21 Haeksim program of the Ministry of 
Education (MOE), by the KOSEF through Center for High Energy Physics (CHEP) 
at Kyungpook National University, 
and by DFG-KOSEF Collaboration program (2000) under the contract 
20005-111-02-2 (KOSEF) and 446 KOR-113/137/0-1 (DFG).

\end{document}